\documentclass[runningheads]{llncs}
\usepackage[T1]{fontenc}
\usepackage{amsmath}
\usepackage{algpseudocode}
\usepackage{algorithm}
\usepackage{algorithmicx}
\usepackage{seqsplit}
\usepackage{fontawesome5}
\usepackage{graphicx}

\newtheorem{pro}{Property}

\newcommand{\nt}{\mathcal{N}_{\mathcal{T}}}
\newcommand{\at}{\mathcal{A}_{\mathcal{T}}}

\newcommand{\dt}{\mathcal{D}_{\mathcal{T}}}
\newcommand{\pt}{\mathcal{P}_{\mathcal{T}}}

\newcommand{\qt}{\mathcal{Q}_\mathcal{T}}

\begin{document}
\title{Solving Time-Dependent Traveling Salesman Problem with Time Windows under Generic Time-Dependent Travel Cost
}
\titlerunning{TSPTW with generic cost function}
%
\author{Duc Minh Vu \faIcon{envelope}\inst{1}\orcidID{0000-0001-5882-3868} \and
Mike Hewitt\inst{2}\orcidID{0000-0002-9786-677X} \and
Duc Duy Vu\inst{3}\orcidID{0000-0002-2657-0445}}
\authorrunning{Vu et al.}
%
\institute{Phenikaa University \& ORLab, Yen Nghia, Ha Dong, Vietnam \\
\faIcon{envelope} Corresponding Author: minh.vuduc@phenikaa-uni.edu.vn \\
 \and
Loyola University Chicago, US\\
\and University of Michigan - Flint, US
}

\maketitle              
\begin{abstract}
In this paper, we present formulations and an exact method to solve the Time Dependent Traveling Salesman Problem with Time Window (TD-TSPTW) under a generic travel cost function where waiting is allowed. A particular case in which the travel cost is a non-decreasing function has been addressed recently. With that assumption, because of both First-In-First-Out property of the travel time function and the non-decreasing property of the travel cost function, we can ignore the possibility of waiting. However, for generic travel cost functions, waiting after visiting some locations can be part of optimal solutions. To handle the general case, we introduce new lower-bound formulations that allow us to ensure the existence of optimal solutions. We adapt the existing algorithm for TD-TSPTW with non-decreasing travel costs to solve the TD-TSPTW with generic travel costs. In the experiment, we evaluate the strength of the proposed lower bound formulations and algorithm by applying them to solve the TD-TSPTW with the total travel time objective. The results indicate that the proposed algorithm is competitive with and even outperforms the state-of-art solver in various benchmark instances.
\end{abstract}

\section{Problem Formulation}
The TD-TSPTW is presented mathematically as follows. We let $(N,A)$ denote a directed graph, wherein the node set $N = \{0, 1, 2, ...,n\}$ includes the depot (node $0$) as well as the set of locations (node $1,\ldots,n$) that must be visited. Associated with each location $i\in N$ is a time window $[e_i,l_i]$ during which the location must be visited. A tourist must visit the city $i$ within its time window. Note that the tourist may arrive at city $i \in N \setminus \{0\}$ before $e_i$, in which case he must wait until the time window opens.  Because of waiting, he does not need to depart immediately after his visit. The time window associated with the depot means that the tour departs the depot at the time of at least $e_0$ and must return to the depot no later than $l_0$.  

We define $A\subseteq N\times N$ as the set of arcs that present travel between locations in N. Associated with each arc $(i,j) \in A$ and time, $t$, at which travel can begin on the arc, is a travel time $\tau_{ij}(t).$ The FIFO property implies that for each arc $(i,j) \in A$ and times $t,t'$ wherein $t \leq t',$ we must have $t + \tau_{ij}(t) \leq t' + \tau_{ij}(t').$ Thus, formally, the vehicle  departs from node $0$ at time $t \geq e_0$, arrives at each city $j \in N$ exactly once within its time window $[e_j,l_j]$ by traveling on arcs in $A,$ and then returns to node $0$ at time $t' \leq l_0.$

We formulate this tour as an integer program defined on a time-expanded network, $\mathcal{D} = (\mathcal{N}, \mathcal{A}),$ with node set $\mathcal{N}$ and arc set $\mathcal{A}$. This formulation is based on the presumption that time may be discretized into a finite set of integer time points. As such, for each node $i \in N, t \in [e_i,l_0]$, $\mathcal{N}$ contains the node $(i,t)$.   $\mathcal{A}$ contains travel arcs of the form $((i,t),(j,t'))$ wherein $i \neq j, (i,j) \in A$, $t \geq e_i, t' = \max\{e_j,t +\tau_{ij}(t)\}$ (the vehicle cannot visit early), and $t' \leq l_j$ (the vehicle cannot arrive late). 
Note that, since waiting is allowed, we can depart from $i$ to $j$ at a time later than the time windows $[e_i,l_i]$ of location $i$. Finally, $\mathcal{A}$ contains arcs of the form $((i,t),(i,t+1))$ presenting waiting at location $i$.  

To formulate the integer program, for each arc $a=((i,t),(j,t')) \in \mathcal{A}$, binary variable $x_a$ represents whether the vehicle travels/waits along that arc. Let $c_a=c_{ij}(t)$ be the non-negative travel cost associated with arc $a$. If $i=j$ and $t'=t+1$, $c_{ii}(t)$ represents the waiting cost for one unit of time period, from time period $t$ to time period $t+1$. Notations $\delta^{+}(i,t)$ and $\delta^{-}(i,t)$ present the set of incoming arcs and the set of outgoing arcs at the node $(i,t)\in \mathcal{N}$. 
The following formulation solves the TD-TSPTW with a generic travel cost function:
 
\begin{equation}
z = \mbox{minimize } \sum_{a \in \mathcal{A}} c_a x_{a} \label{obj:generic}
\end{equation}
subject to
\begin{equation}
\sum_{a=((i,t)(j,t')) \in \mathcal{A}|i\neq j} x_{a} = 1, \;\;\; \forall i \in N, \label{cons:tsptw-arrive-nodes}
\end{equation}
\begin{equation}
\sum_{a \in \delta^{+}(i,t)} x_{a} - \sum_{a \in \delta^{-}(i,t)} x_{a} = 0, \;\;\; \forall (i,t) \in \mathcal{N},i\neq 0,
\label{cons:lb_tsptw-flow-balance}
\end{equation}
\begin{equation}
x_{a} \in \{0,1\}, \;\;\; \forall a \in \mathcal{A}. \label{cons:tsptw-vardef}
\end{equation}

Constraints (\ref{cons:tsptw-arrive-nodes}) ensure that the vehicle arrives at each node exactly one time during its time window. Constraints (\ref{cons:lb_tsptw-flow-balance}) ensure that the vehicle departs every node at which it arrives. Finally, constraints (\ref{cons:tsptw-vardef}) define the decision variables and their domains.  

\section{Literature Review}
Due to the space, we refer \cite{pop2023comprehensive} as a most recent review on TSP in general.
Regarding time-dependent TSPTW literature, \cite{montero2016integer} propose a branch-and-cut algorithm while \cite{ARIGLIANO2018} extend the ideas described in \cite{Cordeau14} by using a branch-and-bound algorithm. \cite{ARIGLIANO2018} show that lower and upper bounds of time-dependent asymmetric TSPTW can be obtained from the optimal solution of a well-defined asymmetric TSPTW. \cite{Boland17b} present the first application of the DDD method to solve the static TSPTW problem. Based on that work, \cite{vu2020dynamic} propose the first DDD method to solve the time-dependent TSPTW problem under the assumption of a non-decreasing travel cost function. The experiment results showed that the algorithm outperformed the state-of-art method for a particular problem of this class, the make-span problem \cite{montero2016integer}. 
\cite{vu2022solving} extend the DDD approach for solving TD-TSPTW to the Time-Dependent Minimum Tour Duration Problem and the Time-Dependent Delivery Man Problem, which, unlike TSPTW,  have a scheduling element.
Regarding how to refine the partial networks, \cite{riedler2019strategies} study path-based refinement strategy and compare results of various refinement strategies applied to TSPTW when using layer graph expansion. 
For a general discussion of Dynamic Discretization Discovery, readers can refer to \cite{boland2019perspectives} as a base source. 

In this paper, we extend the ideas in \cite{vu2020dynamic,vu2022solving} to solve the TD-TSPTW with a generic travel cost function. As far as we know, it is the first research for TD-TSPTW with generic travel costs. We propose lower-bound formulations, and extend the DDD algorithms to find optimal solutions for this generalized problem. 

\section{Partially Time-expanded Network Formulation and Properties}
\label{sec:pte}
To solve TD-TSPTW, we rely on the concept of the {\em partially time-expanded network} \cite{Boland17a,vu2020dynamic}.
A {\em partially time-expanded network}, $\dt = (\nt, \at)$, is derived from a given subset of the timed nodes, $\nt\subseteq \mathcal{N}$. Given $\nt$, the arc set  $\at\subseteq \nt\times \nt$ consists of travel arcs and waiting arcs. A travel arc $((i,t),(j,t'))$, wherein $(i,t) \in \nt$, $(j,t')\in \nt$, $i \neq j,$ and $(i,j) \in A$, models travel between locations $i$ and $j$. We do not allow violations of time windows, so $t' \leq \max\{e_j,t + \tau_{ij}(t)\}$. An arc is \emph{too short} if $t' < \max\{e_j,t + \tau_{ij}(t)\}$. A waiting arc $((i,t)(i,t+1))$ is in $\at$ if both nodes $(i,t)$ and $(i,t+1)$ are in $\nt$. 
For each arc $a=((i,t),(j,t'))\in\at$, we define $\underline{c}_{ij}(t)$ as the travel cost of arc $a$. We set these costs, $\underline{c}_{ij}(t)$, in such a manner that they under-estimate how the cost of such travel is presented in $\mathcal{D}$. Specifically, $\underline{c}_{ij}(t)$ is defined as $\underline{c}_{ij}(t) = \min\{\sum_{h=h'}^{h"-1}c_{ii}(h) + c_{ij}(h") | t \leq h'\leq h" \wedge h" +\tau_{ij}(h") \leq l_j \}$. The underestimated cost $\underline{c}_{ii}(t)$ for a waiting arc $((i,t)(i,t+1))$ is 0 since the waiting cost is taken into account while evaluating underestimated travel cost. 

Given a partially time-expanded $\dt$ that meets Property 1-5, we establish the formulation TD-TSPTW($\dt$) defined by the objective function and constraints (\ref{cons:lb_tsptw-obj}) - (\ref{cons:lb_tsptw-x-defn-pteg}). We optimize this formulation with respect to the cost $\underline{c}_{ij}(t)$.
\begin{equation}
\text{TD-TSPTW($\dt$)}: \min \sum_{a\in \mathcal{A}} \underline{c}_ax_a 
\label{cons:lb_tsptw-obj} 
\end{equation}

\begin{equation}
\sum_{a = ((i,t)(j,t')) \in \at | i\neq j} x_a = 1, \forall i \in N, \label{cons:lb_tsptw-depart-nodes-pteg}
\end{equation}
\begin{equation}
\sum_{a \in \delta^{-}{((i,t))} } x_a - \sum_{a \in \delta^{+}{((i,t))}} x_a = 0, \forall (i,t) \in \nt \label{cons:lb_tsptw-flow-balance-pteg}
\end{equation}
\begin{equation}
x_a \in \{0,1\}, \forall a \in \at. \label{cons:lb_tsptw-x-defn-pteg}
\end{equation}


\begin{pro}
 $\forall i \in N$, both nodes $(i,e_i)$ and $(i, l_i)$ are in $\nt$. \label{prop:open_close_nodes}
\end{pro}	
\begin{pro}
    $\forall (i,t) \in \nt, e_i \leq t. $ \label{prop:what_nodes}
\end{pro}
\begin{pro}
If $(i,t) \in \nt$ and $(i,t+1)\in \nt$, the waiting arc $((i,t)(i,t+1))$ is in $\at$.   	\label{prop:waiting_times}
\end{pro}
\begin{pro}
 Underestimate travel-time arc:	$\forall (i,t) \in \nt$ and  arc $(i,j)\in A$, there is a travel arc of the form $((i,t)(j,t')) \in \at$ if $t + \tau_{ij}(t) \leq l_j$. Furthermore, every  {travel} arc $((i,t),(j,t')) \in \at$ must have either (1) $t + \tau_{ij}(t) < e_j$ and $t' = e_j $, or (2) $e_{j} \leq t' \leq t + \tau_{ij}(t)$. Finally, there is no $(j,t'') \in \mathcal{N}_{\mathcal{T}}$ with $t' < t'' \leq t + \tau_{ij}(t)$. 
 	\label{prop:travel_times}
 \end{pro}
\begin{pro}
	Underestimate travel cost of arc: $\forall ((i,t)(j,t'))\in \at,i\neq j$, the cost $\underline{c}_{ij}(t) = \min\{\sum_{h=h'}^{h"-1}c_{ii}(h) + c_{ij}(h") | t \leq h'\leq h" \wedge h" +\tau_{ij}(h") \leq l_j \}$. Underestimate waiting cost $\underline{c}_{ii}(t)$ takes value 0 for all $i$ and $t$.
	\label{prop:arc_costs}
\end{pro}	

Property 1-5 ensure Lemma \ref{pro:nondecrease}, \ref{pro:nowaiting} and \ref{lmm:lowerbound}. Since the non-decreasing property of $\underline{c}_{ij}(t)$, the algorithm proposed in \cite{vu2020dynamic} will converge to an optimal solution to the TD-TSPTW($\mathcal{D}$) but with the parameterized cost $\underline{c}_{ij}(t)$. Because $\underline{c}_{ij}(t)$ may be not equal ${c}_{ij}(t)$, we may not reach the optimal solution to TD-TSPTW($\mathcal{D}$). Also, using the result of Lemma \ref{pro:nowaiting}, if all optimal solutions to TD-TSPTW($\mathcal{D}$) with the original cost have at least one waiting arc occurring after visiting some cities, then the current lower bound formulation, TD-TSPTW($\mathcal{\dt}$), is not enough to find optimal solutions of TD-TSPTW($\mathcal{D}$). 
\begin{lemma}
     $\underline{c}_{ij}(t)$ is a non-decreasing function of $t$. \label{pro:nondecrease}
\end{lemma}

\begin{lemma} If TD-TSPTW($\dt$) with the parameterized cost $\underline{c}_{ij}(t)$ is feasible, it always has an optimal solution without waiting arcs.  \label{pro:nowaiting}
\end{lemma}


\begin{lemma}
TD-TSPTW($\dt$)  with cost $\underline{c}_{ij}(t)$ is a lower bound of TD-TSPTW($\mathcal{D}$) with cost $c_{ij}(t)$.  \label{lmm:lowerbound}
\end{lemma}

As the first attempt to address the challenges, we extend the lower bound formulation {TD-TSPTW}($\dt$) (\ref{cons:lb_tsptw-obj}) - (\ref{cons:lb_tsptw-x-defn-pteg}) to take into account original travel cost function $c$. Conditions to determine whether we can evaluate an arc with its correct travel cost are presented. Then we introduce new algorithmic ideas and show how to adapt the algorithmic framework presented in \cite{vu2020dynamic} to find optimal solution to TD-TSPTW($\mathcal{D}$) using TD-TSPTW($\dt$). 
Let us state the conditions in which we can evaluate an arc with its correct travel cost. 

\begin{pro} \label{pro:correctcost}
An arc $a=((i,t)(j,t'))\in\dt$ can be evaluated with correct travel cost if and only if two following conditions are met:

\begin{enumerate}
	\item The node $(i,t)$ can be reached by a sequence of correct travel time arcs and waiting arcs from the depot $(0,e_0)$.
	\item The waiting arc $((i,t)(i,t+1))$ is in $\dt$. 
\end{enumerate}
 
\end{pro}


The first condition of Property \ref{pro:correctcost} states that if an arc $a=((i,t)(j,t'))$ can be evaluated with the correct travel cost, then it must be reached from the depot by a sequence of correct travel time arcs (including waiting arcs). With this condition, $t$ is the correct arrival time at the location $i$.
The second condition implies the possibility that we travel from location $i$ at time $t$ to another location with the correct travel cost, or we can wait and travel from $i$ at time at least $t+1$ with underestimate travel cost.  It is used to maintain the non-decreasing property of TD-TSPTW($\dt$) when we update $\dt$.
Now, we present two new formulations satisfying Property 1-6.

\section{New Lower Bound Formulations and Algorithm}
\subsection{Path-arc-based formulation}
We start with a path-arc-based formulation that allows us to evaluate arcs with their correct travel costs by using additional path variables representing paths with correct travel times. Let  $p = ((u_0 = 0,t_0 = e_0)-(u_1,t_1)-(u_2,t_2)-...-(u_m,t_m))$ be a path departing from the depot with correct travel times. We associate with $p$ a binary variable $x_p$ and with cost $c_p = \sum_{i=0}^{m-1}c_{u_{i}u_{i+1}}(t_i)$. Let denote $\pt$ as a set of paths originated from the depot with correct travel times in $\dt$. 
We denote $\pt(i) \subseteq \pt$ as a set of paths visiting the city $i$ and $\delta^{+}_{p}(i,t)$ as a set of paths ending at $(i,t)$. Let  $\delta^{+}(i,t)=\{((j,t')(i,t))\in \at |j\neq i\}$ denote of the set of travel arcs ending at $(i,t)$. Let  $\delta^{+}(i)=\{((j,t')(i,t))\in \at |j\neq i\}$ denote of the set of travel arcs ending at city $(i)$. 
 Let $\at^{=}\subseteq \at$ be the set of arcs with correct travel times, and let $\qt \subseteq \pt \times \at^{=}$ such that if $(p,a) \in \qt$ then $p \oplus a \in \pt$. Here  $p \oplus a =  ((u_0 = 0,t_0=e_0)-(u_1,t_1)-(u_2,t_2)-...-(u_m = i,t_m = t),(j,t'))$, the path obtained by expanding the arc $a$ to the end of $p$ where $p \in \delta^{+}_{p}(i,t)$ and $a =((i,t)(j,t')) \in \at^{=}$.  The path-based formulation TD-TSPTW($\dt,\pt$) is:

\begin{equation}
\text{TD-TSPTW($\dt,\pt$)}: \\ \min \sum_{a \in \at} \underline{c}_ax_a + \sum_{p \in \mathcal{P}_\mathcal{T}} c_px_p    \label{eq:ObjTD}  
\end{equation}

\begin{equation}
x_{a} + \sum_{p \in \mathcal{P}_\mathcal{T} | a\in p} x_{p} \leq 1, \hspace{2mm}  \forall a \in \at^{=}  \label{eq:pathOrArcs}     
\end{equation}

\begin{equation}
x_p +  x_a  \leq 1, \hspace{2mm}  \forall (p,a) \in \mathcal{Q}_\mathcal{T}  \label{eq:pathorarcs1}   \\  
\end{equation}

\begin{equation}
\sum_{a \in \delta^{+}{(i)}}x_a + \sum_{p \in \mathcal{P}_\mathcal{T}(i)} x_p = 1, \hspace{2mm}   \forall i \in N  
\label{eq:tsptw-visitedOnce-extended} 
\end{equation}

\begin{equation}
 \sum_{a \in \delta^{+}{(i,t)}}x_a + \sum_{p \in \delta^{+}_{p}{(i,t)}}x_p - \sum_{a \in \delta^{-}{(i,t)}}x_a = 0, \hspace{2mm} \forall (i,t) \in \nt  \label{eq:tsptw-flow-balance-extended} 
\end{equation}


\begin{equation}
\sum_{((i,t)(i,t+1))\in \at} x_{((i,t)(i,t+1))}=0,\label{eq:nowaiting}  
\end{equation}

\begin{equation}
x_a, x_p \in \{0,1\}, \hspace{2mm}  \forall a\in \at, p\in \mathcal{P}_\mathcal{T}.  \label{eq:x-p-defn}  
\end{equation}

The objective function (\ref{eq:ObjTD}) estimates a lower bound of TD-TSPTW using paths with correct travel times and travel costs plus arcs with under-estimated travel costs.
Suppose $p', p \in \mathcal{P}_\mathcal{T}$ and $p' = p \oplus a$ with some $a\in \at^{=}$ then constraint set (\ref{eq:pathorarcs1}) forces to use $p'$ instead of $p$ and $a$ to ensure that correct cost is used. However, if $p' \notin \mathcal{P}_\mathcal{T}$ and if $p$ and $a$ are selected, we will create $p'$ and add $p'$ to the formulation. Constraint set (\ref{eq:tsptw-visitedOnce-extended}) ensures each city is visited exactly once (e.g. by an arc ending at $(i,t)$ or by a path $p$ passing through node $(i,t)$). Constraint set (\ref{eq:tsptw-flow-balance-extended}) ensures the balance of selected arcs at each node, either by arcs or by paths.  Mathematically, if $x_a = 1$ for any $a \in \delta^{+}{(i,t)}$ in equation (\ref{eq:tsptw-visitedOnce-extended}), then $\sum_{\pt(i,t)}x_p = 0$ and $\sum_{p \in \delta^{+}{(i,t)}}x_p = 0$ in (\ref{eq:tsptw-flow-balance-extended}). So there is an arc $a' \in \delta^{-}{(i,t)}$ with $x_{a'} = 1$, making the node $(i,t)$ balanced. Otherwise,  if $x_a = 0$ for all $a \in  \delta^{+}{(i,t)}$, then $\sum_{\pt(i,t)}x_p = 1$. If $\sum_{p \in \delta^{+}{(i,t)}}x_p = 1$ then again there is an arc $a' \in \delta^{-}{(i,t)}$ with $x_{a'} = 1$, implying the node $(i,t)$ balanced. Otherwise, $\sum_{p \in \delta^{+}{(i,t)}}x_p = 0$, so there is $p \in \mathcal{P}_{\mathcal{T}} \backslash \delta^{+}{(i,t)}$ such that $x_p = 1$. Because $p \notin \delta^{+}{(i,t)}$, there are exactly two arcs with correct travel time of form $((u,h)(i,t))$ and $((i,t)(j,t'))$ in $p$, making $(i,t)$ balance.
Constraint (\ref{eq:nowaiting}), which is used to strengthen the formulation by Lemma 3.2, eliminate waiting arcs with underestimated cost from the optimal solutions, in other words, waiting arcs (with correct travel costs) can only appear in paths in $\pt$. 
Finally, constraint (\ref{eq:x-p-defn}) defines domains of the variables $x$ and $p$.

 %

We have the following results, which say that TSPTW($\dt,\pt$) is always a lower bound of TD-TSPTW($\mathcal{D}$) and explains when we find an optimal solution to TSPTW($\mathcal{D}$). 

\begin{lemma}
TSPTW($\dt,\pt$)  is a lower bound of TSPTW($\mathcal{D}$).
		\label{cor:lower-bound-with-path}
\end{lemma}

\begin{lemma}
		If an optimal solution to TSPTW($\dt,\pt$) is a single tour prescribed by a path (variable), it is an optimal solution to TSPTW($\mathcal{D}$).
	\label{cor:tsptw-optimal-condition-generic}
\end{lemma}


\subsection{Arc-based Formulation}
 
As we can see, there are two major disadvantages of the path-arc-based formulation TD-TSPTW($\dt, \pt$) (\ref{eq:ObjTD})-(\ref{eq:x-p-defn}).
First, the number of paths in $\pt$ can be an exponential number in terms of the number of nodes and arcs in $\dt$, making it impossible to solve the TD-TSPTW($\dt, \pt$) by MIP solver. Second, given a sequence of correct travel time arcs in $\dt$, we cannot evaluate those arcs with correct travel costs unless there is a path including those arcs. We present another modeling approach to solve the two above issues. We associate with each arc $a\in \at$ a variable $z_{a} $ indicating whether this arc can be evaluated with correct travel cost or not. Let $\Delta_{a}=c_{a}-\underline{c}_{a}$ for $\forall a\in\at$. 

The following defines a new relaxation of TD-TSPTW($\mathcal{D}$) that allows evaluating any sequence of arcs with correct travel time from the depot with their correct travel cost when Property 1-6 are met. Let $\nt^{W} = \{(i,t)\in \nt | (i,t+1) \in \nt \}$ be the set of time nodes having waiting arcs, and $\nt^{NW} = \nt \backslash \nt^{W}$ be the set without this property. Let $\lambda^{+}_{(i,t)}$ be the set of \text{correct} travel time/correct travel cost arcs arriving at the timed node $(i,t)$, and $\lambda^{+}_{i}$ be the set of  \text{correct} travel time/correct travel cost arcs arriving at the node $i$.

\begin{equation}
\text{TD-TSPTW($\dt,Z_{\at}$)}: \min\sum_{a\in\at}x_{a}\underline{c}_{a}+\sum_{a\in\at}z_{a}\Delta_{a}\label{eq:z_objLB}
\end{equation}

subject to constraints  (\ref{cons:lb_tsptw-depart-nodes-pteg}), (\ref{cons:lb_tsptw-flow-balance-pteg}) and 
 
\begin{equation}
z_{a}\leq x_{a},\forall a\in\at\label{eq:z_forcingConstraints},
\end{equation}

\begin{equation}
z_{a}=x_{a},\forall a \in \delta^{-}_{(0,e_0)} \text{ if } (0,e_0+1)\in \nt \label{eq:z_atTheDepot},
\end{equation}

 \begin{equation}
z_{a} = 0,\forall (i,t)\in\nt^{NW}, \forall a\in\delta_{(i,t)}^{-} \label{eq:Zforce0},
\end{equation}

\begin{equation}
z_{a}\geq x_{a}+\sum_{a'\in\lambda_{(i,t)}^{+}}z_{a'}-1,\forall (i,t)\in\nt^{W}\backslash (0,e_0), \forall a\in\delta_{(i,t)}^{-} \label{eq:z_connectivity},
\end{equation}

\begin{equation}
\sum_{a\in\lambda_{(i,t)}^{+}}z_{a}\geq\sum_{a\in\delta_{(i,t)}^{-}}z_{a},\forall (i,t)\in\nt\backslash (0,e_0)\label{eq:z_unbalancedZ},
\end{equation}

\begin{equation}
\sum_{a\in \lambda^{+}_{(i,t)}}z_a \geq x_{((i,t)(i,t+1))},\forall (i,t)\in \nt^{W} \backslash(0,e_0) \label{eq:z_noRedundantWaiting}
\end{equation}

\begin{equation}
x_{a},z_a \in\{0,1\},\forall a\in\at\label{eq:z_domainX}.
\end{equation}

 
%
%
%
%


The objective function (\ref{eq:z_objLB})  ensures that if $x_a$ and $z_a$ both take value 1 in a solution to \text{TD-TSPTW($\dt,Z_{\at}$)}, the cost associated to this arc in the objective function is exactly $c_a$. We are going to prove that given a solution $\{\bar{x},\bar{z}\}$ to TD-TSPTW($\dt,Z_{\at}$), for any $a=((i,t)(j,t')) \in \at$, $\bar{z}_a$ takes the value of 1 if and only if Property 6 is met. 

Constraint set (\ref{eq:z_forcingConstraints}) ensures that an arc $a$ is evaluated with its correct travel cost only if this arc is selected in a solution to the formulation. Next, constraint (\ref{eq:z_atTheDepot}) implies that arcs representing departure from $(0,e_0)$ should be evaluated with correct travel costs if waiting arc $((0,e_0),(0,e_0+1)) \in \at $. 
Next, constraint (\ref{eq:Zforce0}) enforces that if there is no waiting possibility at the node $(i,t)$ in the current $\dt$, all arcs outgoing from this node cannot be evaluated with correct travel costs. Otherwise, constraint set (\ref{eq:z_connectivity}) says that if $a=((i,t)(j,t'))\in \at$ is selected and if we can reach the node $(i,t)$ by an arc $a'\in \at^{=}$ which is also evaluated  by its correct travel cost ($\sum_{a'\in\lambda_{(i,t)}^{+}} z_{a'}=1$ or $z_a'=1$ for some $a'$), this constraint forces $a$ to be evaluated by its correct travel cost, or $z_a=1$. 
Constraint (\ref{eq:z_unbalancedZ}) says that if we cannot reach $(i,t)$ by a correct travel time and correct travel cost arc, no outgoing arc of this node can be evaluated with the correct travel cost. Precisely, if $\sum_{a'\in\lambda_{(i,t)}^{+}} z_{a'}=0$, constraint (\ref{eq:z_unbalancedZ}) forces that any outgoing arc $a \in \delta^{-}_{(i,t)}$ will be  evaluated with its underestimate cost since $z_a = 0$ for all $a \in \delta^{-}_{(i,t)}$). 
This constraint also forces $z_{a}=0$ for any arc $a$ in a sub-tour $((u_{0},t_{0}),\,(u_{1},t_{1})\,,...,\,$
$(u_{m},t_{m}),(u_{0},t_{0}))$ where $u_{i}\neq0$ for all $i$. W.r.t, we assume $t_{0}\leq t_{m}$.  Because the too short incoming arc $((u_{m},t_{m}),(u_{0},t_{0}))\notin \lambda^{+}_{(u_0,t_0)}$ of the node $(u_0,t_0)$ is selected, then the left-hand side of constraint (\ref{eq:z_unbalancedZ}) takes the value 0, consequently, $z_a = 0$ for all arcs $a$ of the sub-tour  $((u_{0},t_{0}),\,(u_{1},t_{1}),...,(u_{m},t_{m}),(u_{0},t_{0}))$. Finally, constraint set (\ref{eq:z_domainX}) defines the domain of variables.

In conclusion, the set of constraints (\ref{eq:z_forcingConstraints}) - (\ref{eq:z_domainX}) ensures that if an arc is evaluated with correct travel cost, this arc must belong to the path from the depot node $(0,e_0)$ and all arcs in this path also must be evaluated with correct travel costs.  

\textbf{Aggregation formulation of TD-TSPTW($\dt,Z_{\at}$)} 
Aggregating constrain-ts (\ref{eq:z_atTheDepot}) - (\ref{eq:z_unbalancedZ}) gives us constraints (\ref{eq:za_atTheDepot}) - (\ref{eq:za_unbalancedZ}). $\lambda_{i}^{+}$ and $\delta_{i}^{-}$ present the set of correct-travel-time inbound arcs to city $i$ and set of outbound arcs  from city $i$. The aggregated formulation TD-TSPTW-AGG($\dt,\at$) includes the objective function (\ref{eq:z_objLB}), the constraints  (\ref{cons:lb_tsptw-depart-nodes-pteg}), (\ref{cons:lb_tsptw-flow-balance-pteg}), (\ref{eq:z_forcingConstraints}), 
(\ref{eq:z_domainX}) and the constraints

\begin{equation}
\sum_{a\in\delta_{(0,e_{0})}^{-}}x_{a}=\sum_{a\in\delta_{(0,e_{0})}^{-}}z_{a} \text{ if } (0,e_0+1)\in\nt \label{eq:za_atTheDepot},
\end{equation}
\begin{equation}
\sum_{(i,t)\in \nt^{NW}}\sum_{a\in\delta_{(i,t)}^{-}}z_{a}=0 \label{eq:za_noforceZ},
\end{equation}
\begin{equation}
\sum_{a\in\delta_{(i,t)}^{-}}z_{a}\geq\sum_{a\in\delta_{(i,t)}^{-}}x_{a}+\sum_{a\in\lambda_{(i,t)}^{+}}z_{a}-1,\forall(i,t)\in\nt^{W}\backslash \{0,e_0\} \label{eq:za_connectivity},
\end{equation}
\begin{equation}
\sum_{a\in\lambda_{i}^{+}}z_{a}\geq\sum_{a\in\delta_{i}^{-}}z_{a},\forall i \in N\backslash 0\label{eq:za_unbalancedZ}.
\end{equation}
Actually, while TD-TSPTW-AGG($\dt,Z_{\at}$) is the aggregated version of TD-TSPTW($\dt,Z_{\at}$), we can prove that they are equivalent (see Appendix).
Similar to the results of path-based formulation, we have:
 \begin{lemma}
	If $\dt$ satisfies Properties \ref{prop:open_close_nodes} - \ref{prop:arc_costs}, then TD-TSPTW($\dt,Z_{\at}$) (or TD-TSPTW-AGG($\dt,Z_{\at}$), respectively) is a relaxation of TD-TSPTW($\mathcal{D}$).
	\label{lem:za_td-tsptw-lb}
\end{lemma}
\begin{lemma}
 An optimal solution to TD-TSPTW($\dt,\at$) (or TD-TSPTW-AGG ($\dt,Z_{\at}$), respectively) that has no too short arcs defines an optimal solution to TD-TSPTW($\mathcal{D}$)
\label{lem:za_non-decreasing-stopping}
\end{lemma}
\subsection{Algorithm to solve TD-TSPTW($\dt$)}
\begin{algorithm}[t]
	\caption{\textsc{DDD-TD-TSPTW}} \label{alg:solve_cont_ts_enhanced}
	\begin{algorithmic}[1]
		\Require TD-TSPTW instance $(N,A)$, $e$, $l$, $\tau$ and $c$, and optimality tolerance $\epsilon$
		\State Perform preprocessing, updating $A$, $e$ and $l$.
		\State Create a partially time-expanded network $\dt$.
		\State Set $\pt \leftarrow \emptyset$, (or $\mathcal{A}_{T}^{=} \leftarrow \text{the set of correct travel time arcs in } \at)$
		\State Set $S \leftarrow \emptyset$  
		\While{not solved}
		\State Set $\bar{S}\leftarrow \emptyset$ 
 		\State Solve primal heuristics, TD-TSPTW($\dt^1$) and TD-TSPTW($\dt^2$), with underestimate cost $\underline{c}$,  harvest integer solutions and add to  $\bar S$. \label{step:primalheuristics}
		
		%
		\State Solve TD-TSPTW($\dt$),   harvest integer solutions, $\bar S$, and lower bound, $z$.
        \For{$s \in \bar S$}
		\State Let $s'$ be a copy of $s$ without waiting arcs. \label{step:copy}
		\If{$ s'$ can be converted to a feasible solution to TD-TSPTW($\mathcal{D}$)} \label{step:check}
		\State Solve R-TD-TSPTW($\mathcal{D},s'$) to find the best tour using travel arcs in $s'$. \label{step:solveRestrict}
		\State Update S with the solution returned by R-TD-TSPTW($\mathcal{D},s'$).
		\State Add a cut to exclude all copies of $s'$ in TD-TSPTW($\dt$).\label{step:excludefeasible}
		\Else
		\State Add cuts corresponding to sub-tours and infeasible paths in $s'$ to TD-TSPTW($\dt$) \label{step:addcuts}
		\EndIf
 		\State Update $\dt$, (and $\pt$), and TD-TSPTW($\dt$) by lengthening arcs in $s$. \label{step:updatenetwork}
		\EndFor
		\State Compute gap $\delta$ between the best solution in $S$ and lower bound, $z$.
		\If{$\delta \leq \epsilon$}
		\State Stop: best solution in $S$ is $\epsilon-$optimal for TSPTW. \label{alg:stop_condition}
		\EndIf
		\EndWhile
	\end{algorithmic}
\end{algorithm} 

Algorithm \ref{alg:solve_cont_ts_enhanced} shows how we solve the TD-TSPTW($\mathcal{D}$) using proposed lower bound formulations.  Readers can refer to \cite{Boland17b,vu2020dynamic,vu2022solving} for additional reference of the components of the algorithm. 
In Algorithm \ref{alg:solve_cont_ts_enhanced}, TD-TSPTW($\dt$) refers to one of the lower bound formulations presented in Section 4. While it shares the main steps with the one mentioned in \cite{vu2020dynamic,vu2022solving}, there are differences. First, to check the feasibility of a solution $s$ found by lower bound formulations, we check with the solution $s'$ obtained from $s$ excluding all waiting arcs (Line \ref{step:copy}-\ref{step:check}). It ensures that if $s'$ is infeasible, $s$ is always infeasible. If $s'$ is feasible, we solve R-TD-TSPTW($D,s'$) to find the best tour using only travel arcs in $s'$ (e.g. by adding waiting arcs to $s'$,  Line \ref{step:solveRestrict}). R-TD-TSPTW($D,s'$)  is a restricted formulation of TD-TSPTW($D$) where only travel arcs in $s'$ can be selected. If R-TD-TSPTW($D,s'$) is solved to optimality, we add a cut to exclude all copies of $s'$ from TD-TSPTW($\mathcal{\dt}$). If $s'$ is infeasible, we add cuts corresponding to sub-tours and infeasible paths extracted from $s'$ to TD-TSPTW($\dt$). The two primal heuristics in \cite{vu2020dynamic,vu2022solving} are used to help find feasible solutions. We exclude all waiting arcs when solving those two primal heuristics (Line \ref{step:primalheuristics}). Arcs in those primal heuristics are evaluated with under-estimate cost $\underline{c}_{ij}(t)$ and with basic formulation (5)-(8). We add cuts to exclude all tours corresponding to feasible solutions in $S$ before solving those primal problems to force finding new feasible solutions. When updating $\dt$ (Line \ref{step:updatenetwork}), given an arc $((i,t)(j,t'))$ to lengthen, we also add the node $(i,t+1)$ to $\dt$ to introduce waiting opportunities at $(i,t)$ if $(i,t+1) \notin \dt$. To maintain Property 1-5, when a new node $(i,t)$ is added to $\dt$, we also add arc $((i,t)(i,t+1))$ (or $((i,t-1)(i,t))$) if node $(i,t+1)$ (or $(i,t-1)$) existed in $\dt$. When the path-arc-based formulation is used, Algorithm \ref{alg:add-path} (see Appendix) is employed to update $\pt$.


		
		
 
		

\section{Experiments}
The algorithm is implemented in C++ using Gurobi 8.0 as the MIP solver. All experiments were run on a workstation with an Intel(R) Xeon (R) CPU E5-4610 v2 2.30GHz processor running Ubuntu 14.04.3. One of the stopping conditions was a provable optimality gap of $\epsilon = 10^{-2}$. Two sets named "Set 1" and "Set w100" from \cite{vu2020dynamic} are used. Each set has 960 instances from 15 to 40 nodes with up to 73 travel time profiles. We set the cost $c_{ij}(t)$ be $\tau_{ij}(t)$, the travel time between $i$ and $j$ at time point $t$. We assess two points:
\begin{enumerate}
    \item We compare the path-arc-based formulation (Path), the arc-based formulation (Z), and the aggregated arc-based formulation (Z-Agg) and the corresponding algorithms based on the number of instances solved.
    \item We compare the strongest algorithm and formulation and the state-of-the-art solver, Gurobi,  solving the problem with the full-time-expanded network formulation.
\end{enumerate}

First, Table \ref{tbl:compareFormulations} reports the number of solved instances using the Path, Z, and Z-Agg formulation. In this experiment, we consider a setting in which waiting at site $i$ after visiting $i$ is not allowed, so $c_{ii}(t) = \infty$, or a very high value. Maximum running time for this setting is 1 hour. This can happen for time-dependent scheduling problems where all jobs (cities) are performed without stopping.  As we expect, the Z-Agg formulation is the most efficient and competitive formulation, while the Path formulation is the worst one. Using the Z-Agg formulation, we can solve 873 and 871 instances of Set 1 and Set w100 to optimality. It means the proposed algorithm is able to solve this particular variant. 
\begin{table}[H]
\begin{center}
\caption{Number of instances solved optimally by each formulation.}
\begin{tabular}{|c|c|c|c|c|c|c|}
\hline 
 & \multicolumn{3}{c|}{Set 1} & \multicolumn{3}{c|}{Set w100}\tabularnewline
\hline 
$n$ & Path & Z & Z-Agg & Path & Z & Z-Agg\tabularnewline
\hline 
15 & 240 & 240 & 240 & 240 & 240 & 240\tabularnewline

20 & 239 & 239 & 240 & 226 & 228 & 231\tabularnewline

30 & 214 & 218 & 224 & 192 & 230 & 232\tabularnewline

40 & 147 & 149 & 169 & 115 & 167 & 188\tabularnewline
\hline 
\end{tabular}
\par\end{center}
\label{tbl:compareFormulations}
\vspace{-4mm}
\end{table}
Second, we consider the setting in which $c_{ii}(t) = 0$. This setting is harder because of the larger solution space, so we let 2 hours of execution. We observe that while Gurobi can efficiently solve instances of Set w100, it struggles to find feasible solutions to instances of Set 1. Given two hours of computation time, it can only find feasible solutions to 596 over 960 instances of Set 1 (Table \ref{tbl:feasibleGurobi}), while it is able to find feasible solutions to all instances of Set w100. Technically, while having the same number of nodes, instances of Set 1 have wider time windows, making the complete networks larger and harder to solve. The proposed algorithm finds feasible solutions for all instances.
\begin{table}[H]
\begin{centering}
\caption{Feasible solutions: Gurobi versus Z-Agg (Set 1)}
\begin{tabular}{|c|c|c|c|c|}
\hline 
$n$ & 15 & 20 & 30 & 40\tabularnewline
\hline 
\hline 
Gurobi & 239 & 228 & 115 & 14\tabularnewline
\hline 
Z-Agg & 240 & 240 & 240 & 240\tabularnewline
\hline 
\end{tabular}
\par\end{centering}
\label{tbl:feasibleGurobi}
\end{table}
 
Finally, Table \ref{tbl:optimal} compares the number of $\epsilon$-optimal solutions that Gurobi and the proposed algorithm find for instances of Set 1. Gurobi can prove optimality for 396 instances, while the proposed method with Z-Agg formulation can solve 712 instances. The average gap of unsolved instances is 7.37\% (Gurobi) and 2.94\% (the proposed algorithm). These preliminary results show that the proposed algorithm and formulations are promising for solving time-dependent TSPTW instances.
\begin{table}[H]
\begin{centering}
\caption{Optimal solutions: Gurobi versus Z-Agg (Set 1)}
\begin{tabular}{|c|c|c|c|c|}
\hline 
$n$ & 15 & 20 & 30 & 40\tabularnewline
\hline 
\hline 
Gurobi & 218 & 145 & 29 & 4\tabularnewline
\hline 
Z-Agg & 228 & 199 & 157 & 128\tabularnewline
\hline 
\end{tabular}
\label{tbl:optimal}
\par\end{centering}
\end{table}
To conclude, the three lower bound formulations can be used to solve the time-dependent TD-TSPTW in which the aggregated formulation TD-TSPTW-AGG($\dt,\at$) is the most effective one. It has fewer constraints than TD-TSPTW($\dt,\at$), and therefore, makes it easier to solve with mixed-integer programming solvers.  The proposed algorithm with the aggregated formulation performs better than the solver over benchmark instances.

\section{Conclusion}

In this paper, we study a generalized version of the time-dependent traveling salesman problem where travel cost is modeled as a generic function. We present three lower-bound formulations based on path and arc variables, and we introduce iterative exact algorithms based on the dynamic discrete discovery approach to solve the problems. The experiment results confirm the advantages of the proposed method over the state-of-art solvers when solving small- and medium-sized instances. As a final note, our ongoing research shows that we can apply the proposed method in this paper to solve variants of the TSPTW including those with soft time windows. It again confirms the strength and innovation of our proposed method.
\section*{Acknowledgement}
The work has been carried
out partly at the Vietnam Institute for Advanced Study in Mathematics (VIASM). The corresponding author (Duc Minh Vu) would like to thank VIASM for its hospitality and financial support for his visit in 2023.

\bibliographystyle{plain}
\bibliography{main}

\section{Appendix}
\subsection*{Algorithm \textsc{Add-Paths}}
Algorithm \textsc{Add-Paths} presents how we maintain and update the set of paths $\pt$. We start with an empty set $\pt$. Let $c= ((u_0 = 0,t_0 = e_{u_0}),(u_1,t_1),$ $...,(u_m = 0,t_m))$ be a tour prescribing a sequence of arcs departing from the depot. For each sub-path $c_i = ((u_0 = 0,t_0 = e_{u_0}),(u_1,t_1),...,(u_i,t_i))$, we generate a path variable corresponding to this sub-path and add this path to $\pt$. We also generate new waiting opportunities at each node to maintain Property \ref{pro:correctcost}.    

\begin{algorithm}
	\caption{\textsc{Add-Paths}(c)} \label{alg:add-path}
	\begin{algorithmic}[1]
		\Require a tour c = $((u_0 = 0,t_0 = e_{u_0}),(u_1,t_1),...,(u_m = 0,t_m))$.
		\State $p \leftarrow \emptyset$
		\State $t_0' \leftarrow t_0$
		\For{$k \leftarrow 1$ to $m$ }
		
		\If{$(u_{k-1},t_{k-1}'+1) \notin \nt$}
		\State Add $(u_{k-1},t_{k-1}'+1)$ to $\nt$ and update $\dt$. 
		\EndIf
		\If{$p \oplus ((u_{k-1},t_{k-1}'),(u_{k-1},t_{k-1}'+1)) \notin \pt $} 
		\State  Add $p \oplus ((u_{k-1},t_{k-1}'),(u_{k-1},t_{k-1}'+1)) $ to $\pt$.  
		\EndIf
		
		\State $t_k' \leftarrow \max(e_{u_k}, t_{k-1}' +\tau_{u_{k-1}u_k}(t_{k-1}'))$	 
 
		\If{$t_{k}' > l_{u_{k}}$}
		\State Break;
		\EndIf
		
		\State $p \leftarrow p \oplus ((u_{k-1},t_{k-1}'),(u_k,t_k'))$
		\State If $p \notin \pt $, add $p$ to $\pt$.
		\EndFor
	\end{algorithmic}
\end{algorithm}

\subsection*{Proof of Lemma \ref{pro:nondecrease}}
\proof{} The result is obtained directly from the definition of $\underline{c}_{ij}(t)$ and the FIFO property of travel time function $\tau_{ij}(t)$.
 
\subsection*{Proof of Lemma \ref{pro:nowaiting}}
\proof{} It is because of the non-decreasing property of travel cost function $\underline{c}_{ij}(t)$ and because underestimate waiting cost takes the value 0 for all $i$ and $t$, there is an optimal solution to TD-TSPTW($\dt$) without waiting arcs.


\subsection*{Proof of Lemma \ref{lmm:lowerbound}}
As proved in \cite{vu2020dynamic},  TD-TSPTW($\dt$) with cost $\underline{c}_{ij}(t)$ is a lower bound of TD-TSPTW($\mathcal{D}$) with cost $\underline{c}_{ij}(t)$. Also, TD-TSPTW($\mathcal{D}$) with cost $\underline{c}_{ij}(t)$ is a lower bound of TD-TSPTW($\mathcal{D}$) with cost ${c}_{ij}(t)$. Therefore,  TD-TSPTW($\dt$) with cost $\underline{c}_{ij}(t)$ is a lower bound of TD-TSPTW($\mathcal{D}$) with cost ${c}_{ij}(t)$, the lemma is proved.

\subsection*{Proof of Lemma \ref{cor:lower-bound-with-path}}
\proof{}

We begin the proof with an observation that, given a feasible solution $p = ((u_0 = 0,t_0 = e_0),(u_1,t_1),...,(u_{m} = 0, t_{m}))$ to TD-TSPTW($\mathcal{D}$), we can decompose $p$ into two sub - paths as $p = p_1 \oplus p_2$ where $p_1 =  ((u_0 = 0,t_0 = e_0),(u_1,t_1) ... (u_{l},t_{l})) \in \pt$ and $p_2  = ((u_{l},t_{l}),(u_{l+1},t_{l+1}), ..., (u_{m}= 0,t_{m}=e_0)) \in \mathcal{D}$.  
For the path $p_2 = ((u_{l},t_{l}),(u_{l+1},t_{l+1}), ..., (u_{m}= 0,t_{m}=e_0)) \in \mathcal{D}$, there is a copy $p_2' = ((u_l' = u_l,t_l' = t_l),(u_{l+1}',t_{l+1}'), ..., (u_{m'}= 0,t_{m'}'=e_0)) \in \dt $ with $t_i' \leq t_j$ if $u_{i}'=u_j$. Note that, $p_2$ may have waiting arcs while $p_2'$ does not.  The cost $\underline{c}_{p_2'} = \sum_{i=l}^{m'} \underline{c}_{u_iu_{i+1}}(t_i')$ does not exceed the cost $c_{p_2} = \sum_{i=l}^{m} {c}_{u_iu_{i+1}}(t_i)$ by the definition of  $\underline{c}$.

The path $p_1 \oplus p_2'$ defines a feasible solution to TSPTW($\dt,\pt$). Since the cost of $p_2'$ cannot exceed the cost of $p_2$, therefore, TSPTW($\dt,\pt$) is a lower bound of TSPTW($\mathcal{D}$). \qed

%

\endproof
\subsection*{Proof of Lemma \ref{cor:tsptw-optimal-condition-generic}}
\proof Because the optimal tour to  TSPTW($\dt,\pt$) is prescribed by a path variable, its value is then exactly the true travel cost induced by the tour in the full-time-expanded network. So it is optimal to TSPTW($\mathcal{D}$). \qed
\endproof

\subsection*{Proof of Lemma \ref{lem:za_td-tsptw-lb}}

\proof Suppose $p= ((u_0 = 0, t_0=e_0),  (u_1,t_1),$  $\ldots, (u_{m}=0,t_{m}=e_0)) \in \mathcal{D}$ is a feasible solution to TD-TSPTW($\mathcal{D}$). Suppose $l\leq m$ be the largest value such that the sub-path $p_1= ((u_0 = 0, t_0=e_0),  (u_1,t_1),...,(u_{l},t_{l}))\in \dt$ and $(u_k,t_k+1)\in \nt$ for all $k=0..l$ (waiting condition). Let $p = p_1 \oplus p_2$ where $p_2$ is the remaining of $p$ after excluding $p_1$.  
Let $p_2' = ((u_{l},t_{l}' = t_{l}),(u_{l+1},t_{l+1}'),...,(u_{r}'=0,t_{r}')) \in \dt$ be the copy of $p_2$ without waiting arcs. 
Following the definition, $p'= p_1 \oplus p_2'$ is a solution to TD-TSPTW($\dt,Z_{\at}$) where $p'$ shares the longest prefix with $p$ with the waiting condition. Also, it means that, for each solution $p$ to the original problem, we have exactly one copy solution $p'$ to the lower bound problem. The two solutions have the same prefix $p_1$; and we have exactly one copy $p'$ since there is only one copy $p_2'$ of $p_2$ that contains only travel arcs.

Let $p_2' = ((u_l,t_l)(u_{l+1},t_{l+1}')) \oplus p_2^{"}$ where $p_2^{"} = ((u_{l+1},t_{l+1}'),...,(u_{r}'=0,t_{r}')))$. We have the cost of $p_2'$  evaluated by the objective function to TD-TSPTW($\dt, Z_{\at}$)   is $c_{u_{l}u_{l+1}}(t_{l}) + \underline{c}(p_2'')$. It is since we evaluate the cost of the arc $((u_l,t_l),(u_{l+1},t_{l+1}'))$  with correct travel cost because we reach the node $(u_l,t_l)$ by a sequence of arcs with correct travel time and the node $(u_l,t_l+1)$ is in $\dt$. Therefore, the cost of $p'$ with respecting the relaxation TD-TSPTW($\dt,Z_{\at}$)  is $c(p_1) + c_{u_{l}u_{l+1}}(t_{l}) + \underline{c}(p_2")$. Let $p_2''' = ((u_q,t_q),...,(u_m,t_m))$ is the longest suffix of $p \in \mathcal{D}$ where $q$ is the smallest index that $u_q = u_{l+1}$. Since $q\geq l+1$, therefore, $t_q \geq t_{l+1}'$. The cost $c_{p_2'''}$ is at least as large as the cost $\underline{c}_{p"}$. Therefore, the cost of $p'$ evaluated by the lower bound formulation cannot exceed the cost $p$. 
In conclusion,  TD-TSPTW($\dt,Z_{\at}$) is a lower bound to the TD-TSPTW($\mathcal{D}$). \qed

\subsection*{Proof of Lemma \ref{lem:za_non-decreasing-stopping}}
\proof Since $s$ consists of correct travel time arcs, it is also a solution to  TD-TSPTW($\mathcal{D}$). Since  TD-TSPTW($\dt,\at$) is a lower bound of TD-TSPTW($\mathcal{D}$), and we evaluate $s$ with correct travel costs, therefore, it defines a feasible solution to the original problem. \qed

\begin{lemma}
The value of \text{TD-TSPTW-AGG($\dt,Z_{\at}$)} is equal to the value of \text{TD-TSPTW($\dt,Z_{\at}$)}.
\label{lem:equivalent}
\end{lemma}

\proof{} 

To prove this result, we will prove that we can obtain all constraints of the disaggregated formulation from the constraints of the aggregated formulation.
 
First, Constraint (\ref{eq:z_forcingConstraints}) and (\ref{eq:za_atTheDepot}) ensure $x_a=z_a$ for $a=((0,e_0)(i,t))$, which is Constraint (\ref{eq:z_atTheDepot}). Constraint (\ref{eq:za_noforceZ}) also ensures constraint (\ref{eq:Zforce0}) 
happens. Next, suppose that $0\leq z_{a} < x_{a}+\sum_{a'\in\lambda_{(i,t)}^{+}}z_{a'}-1\leq 1$, for some  $a\in\delta_{(i,t)}^{-}$ or constraint (\ref{eq:z_connectivity}) violates at node $(i,t)$ for an outgoing arc $a$ while Constraint (\ref{eq:za_connectivity}) is met.  It means that $z_a=0$, and $ x_{a}= \sum_{a'\in\lambda_{(i,t)}^{+}}z_{a'} = 1$. Therefore, we must have $x_{\bar{a}} = z_{\bar{a}} = 1$ for some $\bar{a} \in \delta_{(i,t)}^{-} \backslash a$ to force (\ref{eq:za_connectivity}) happen. Consequently, we have at least two outgoing arcs from the node $(i,t)$, violating Constraint (\ref{cons:lb_tsptw-depart-nodes-pteg}). So, constraint (\ref{eq:z_connectivity}) is also valid. 


Finally, suppose that $0\leq \sum_{a\in\lambda_{(i,t)}^{+}}z_{a}<\sum_{a\in\delta_{(i,t)}^{-}}z_{a}\leq 1$ for a node $(i,t)$ or constraint (\ref{eq:z_unbalancedZ}) violates at the node $(i,t)$ while Constraint (\ref{eq:za_unbalancedZ}) is met for a location $i$. It leads to $\sum_{a\in\lambda_{(i,t)}^{+}}z_{a}=0$, and $\sum_{a\in\delta_{(i,t)}^{-}}z_{a}=1$. Suppose $a_1 = ((i_1,t_1),(i,t)) \in \lambda_{(i,t)}^{+}$ and $a_2 = ((i,t),(i_2,t_2)) \in \delta_{(i,t)}^{-}$ with $z_{a_1} = 0$ and $z_{a_2} = 1$ be two incoming and outgoing arcs associated to the node $(i,t)$. Because $z_{a_1} = 0$, so $a_2$ must be an outgoing travel arc from $(i,t)$ because of the constraint (\ref{eq:z_noRedundantWaiting}). If $a_1$ is a travel arc, or $i_1 \neq i$, so to satisfy the Constraint (\ref{eq:za_unbalancedZ}) with respect to location $i$, we need to visit $i$ at least two times, which violates the visiting-once condition of TSPTW.
If $a_1$ is a waiting arc, or $a_1 = ((i,t-1),(i,t))$, in this case, constraint (\ref{eq:z_noRedundantWaiting}) ensures that $\sum_{a\in\lambda^{+}_{(i,t-1)}} z_a = 1$, otherwise, $x_{(i,t-1)(i,t)} = 0$ and $a_1$ is not selected. However, since $a_1$ is waiting arc and $x_{(i,t-1)(i,t)} = 1$, constraint (\ref{eq:za_connectivity}) forces $z_{(i,t-1)(i,t)}$ or $ z_{a_1} $ to be 1, contradictory to the assumption that $\sum_{a\in\lambda_{(i,t)}^{+}}z_{a}=0$. We can conclude that the constraint set (\ref{eq:z_unbalancedZ}) is met when constraint set (\ref{eq:za_unbalancedZ}) is met.

So all constraints of \text{TD-TSPTW($\dt,Z_{\at}$)} can be obtained from constraints \text{TD-TSPTW-AGG($\dt,Z_{\at}$)}, therefore, two formulations are equivalent. \qed

Finally, the following result ensures the convergence of the lower bound values when the partially time-expanded networks are updated. 
\begin{lemma}
Given that $\nt \subseteq \nt'$ and $\pt\subseteq \pt'$, the value of
TD-TSPTW($\dt'$, $\pt'$) (or \text{TD-TSPTW($\dt',Z_{\at'}$)}, \text{TD-TSPTW-AGG($\dt',Z_{\at'}$)}) is at least as large as the value of
\text{TD-TSPTW($\dt,\pt$)} (or \text{TD-TSPTW($\dt,Z_{\at}$)},  TD-TSPTW-AGG($\dt,$ 
$ Z_{\at}$))
\label{lem:nondecreasing}
\end{lemma}
\proof{} The proof for this result is based on an observation that if $\nt \subseteq \nt'$, we can map any path $p'$ in $\dt'$ to a path $p$ in $\dt$ with cost $c_p' \geq c_p$.  See \cite{vu2020dynamic} for the proof of the observation.

\end{document}